\documentclass[journal,twoside,final]{IEEEtran}

\usepackage{amsmath}
\usepackage{amssymb}
\usepackage{graphicx}
\usepackage[caption=false,font=footnotesize]{subfig}
\usepackage{cite}

\begin{document}

\title{Pointing Error Analysis of Optically Pre-Amplified Pulse Position Modulation Receivers}
\author{Konstantinos~Yiannopoulos, Nikolaos~C.~Sagias,~\IEEEmembership{Senior~Member,~IEEE}, and~Anthony~C.~Boucouvalas,~\IEEEmembership{Fellow,~IEEE}
\thanks{K. Yiannopoulos, N. C. Sagias and A. C. Boucouvalas are with the Department of Informatics and Telecommunications, University of the Peloponnese, Akadimaikou G. K. Vlachou Street, GR-221 31 Tripoli, Greece (email: {\tt kyianno@uop.gr, nsagias@uop.gr, acb@uop.gr}).}}

\maketitle

\begin{abstract}
We present analytical results on the effect of pointing errors on the average bit-error probability (ABEP) of optically pre-amplified pulse-position modulation (PPM) receivers. The results show that the beam width plays a key role in the ABEP and that a significant power penalty is introduced by utilizing sub-optimal widths, especially when pointing errors incorporate a jitter component. We also present the optimisation the beam width for a number of pointing error scenarios and show that increasing beam widths are required as additional signal energy becomes available. The optimal beam width is also affected by the PPM modulation order and the optical noise modes, with more energy efficient (higher modulation order and lower noise) systems allowing for broader beams.
\end{abstract}

\begin{IEEEkeywords}
Optical wireless communications, pointing errors, optical amplifiers, pulse position modulation.
\end{IEEEkeywords}

\section{Introduction}\label{sect:intro}  
Optical amplification has been extensively studied for the compensation of the losses that are introduced by the transmission of optical signals through the atmosphere or over ultra-long link lengths, as is typically the case in space communications \cite{J:Arnon05}. Optical amplification imparts a significant improvement on the receiver sensitivity, which can be further increased by utilizing a power efficient modulation format like PPM, and the combination of amplification and PPM was recently demonstrated in a high-speed link with the Moon \cite{J:Boroson14}. The communication over optical beams, however, can be disrupted due to pointing errors, since the transmitted beam width is typically very narrow and any misalignment between the transmitter and the receiver introduces a power loss that negatively affects the ABEP link performance.

In this work, we present novel analytical results on the ABEP performance of an optically pre-amplified PPM receiver in the presence of pointing errors. The results show that the received beam width plays a key role in the performance and must be optimised so as to minimize the ABEP. The results also show that the optimal beam width is strongly dependent on the total optical energy at the receiver location. For low beam energies, the optimal width is practically constant irrespective of the modulation order and noise modes, and is comparable to the static misalignment. As more energy becomes available, the beam should be gradually expanded so as to also mitigate the effect of random misalignments.

\IEEEpubidadjcol

\section{System Model}\label{sect:sysmod}  
\begin{figure}[!t]
\centering
\subfloat{\includegraphics[keepaspectratio,width=\columnwidth]{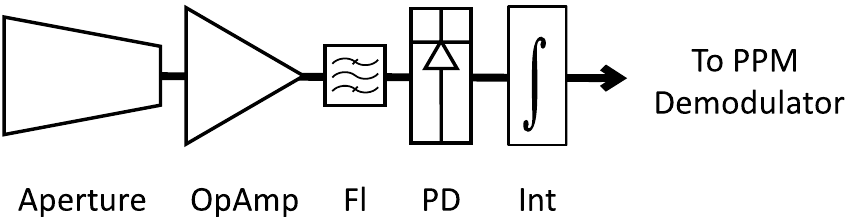}}
\caption{Optically pre-amplified PPM receiver. OpAmp: Optical amplifier; Fl: Optical Filter; PD: Photodiode; Int: Time-slot integrator.}
\label{fig:setup}
\end{figure}

We consider an optical-wireless PPM receiver whose structure is detailed in Fig.~\ref{fig:setup}. Light is collected from an optical aperture and is amplified prior to detection, with the amplifier providing a gain equal to $G$ and adding optical noise with a spectral density of $N_{0}=n_{sp} \, h \, f \, (G-1)$,  where $h \, f$ is the photon energy and $n_{sp}$ is the spontaneous emission factor of the amplifier. An optical filter is utilized to reject amplifier noise and the optical signal is then converted to electrical on a photodiode. The photodiode output is integrated over the duration of a PPM time-slot and soft decision decoding is utilized to identify the slot with the highest signal and decode the symbol.

We now extend a previous work that addressed the impact of fading on the ABEP of the PPM receiver \cite{J:Yiannopoulos20} to model the impact of pointing errors. Following our previous work, the ABEP of the optically pre-amplified PPM receiver is given by

\begin{equation}\label{Eq:chisqSerrorprobPPM}
\begin{split}
\overline{P_{e}} & = \frac{Q}{2\,(Q-1)} \, \sum_{q=1}^{Q-1} \, \binom{Q-1}{q} \, (-1)^{q+1} \\
 & \times \, \sum_{n=0}^{q\,(M-1)}  \sum_{i=n}^{q\,(M-1)} \, \binom{i+M-1}{n+M-1} \, \frac{c_{i}^{q}}{(1+q)^{i+M}} \, \frac{w(n)}{q^n} \,,
\end{split}
\end{equation}
with
\begin{equation}\label{Eq:chisqSerrorprobPPMavg}
w(n)  = \frac{z_{q}^{n}}{n!} \, \int_{0}^{\infty} t^n \,  f_t(t) \, \exp \left( - z_{q} \, t \right) \, dt \, ,
\end{equation}
where $Q$ is the PPM modulation order, $M$ are the optical noise modes, $c_{i}^{q}$ are constants, $z_{q} = \frac{q}{1+q}\, E_{b}/N_{0} \, \log_{2}(Q)$, $E_{b}~=~G \, E_{in}$, $E_{in}$ is the total energy (per bit) of the beam at the receiver location and $t$ denotes the fraction of the total energy that is collected by the aperture. $t$ depends on the instantaneous misalignment between the beam and aperture centers, and for the purposes of this work it is treated as a random variable (RV) with a probability density function $f_t(t)$.

\begin{figure*}[!t]
\centering
\subfloat{\includegraphics[keepaspectratio,width=0.9\columnwidth]{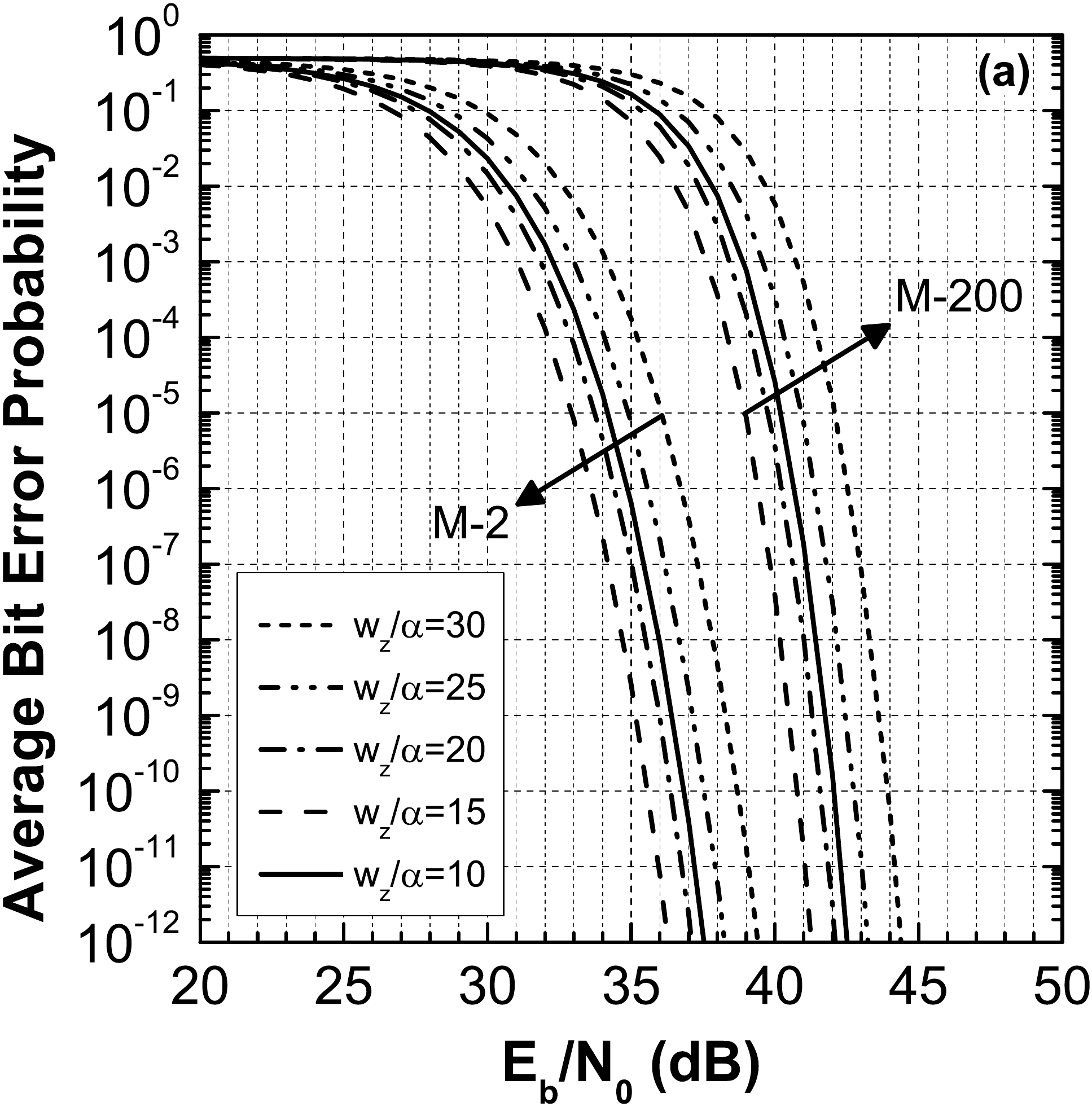}
\label{Fig:ABEPstatic}}
\hfil
\subfloat{\includegraphics[keepaspectratio,width=0.9\columnwidth]{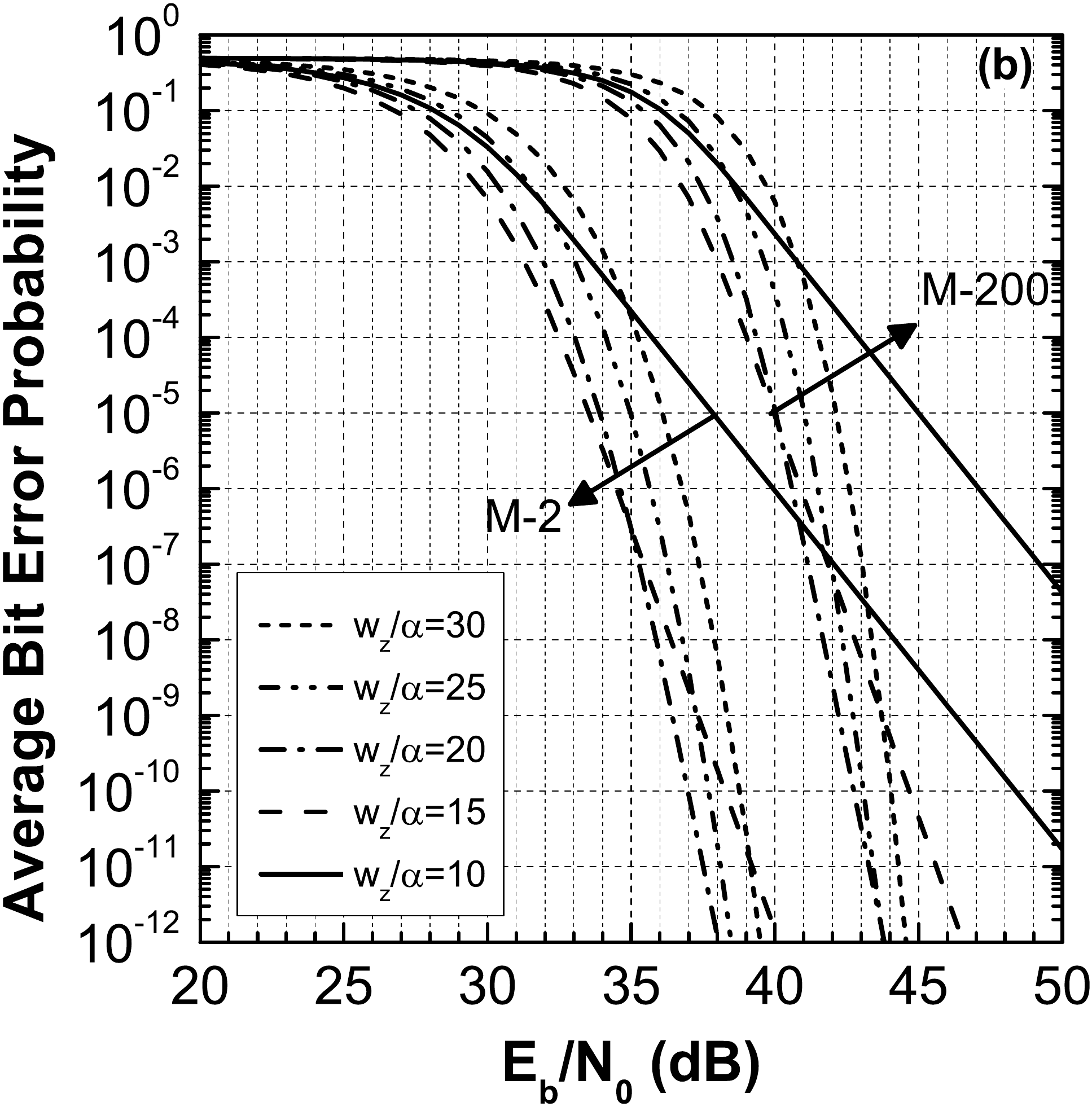}
\label{Fig:ABEPjitter}}
\caption{16-PPM ABEP in the presence of pointing errors for (a) static misalignment, and (b) misalignment with jitter.}
\label{fig:ABEP}
\end{figure*}

\section{Mathematical Analysis}\label{sect:math}  
Assuming that the misalignments in the $x$ and $y$ axis are Gaussian RVs with mean values equal to $\mu_{x}$ and $\mu_{y}$ and variances (jitter) equal to $\sigma_{x}$ and $\sigma_{y}$, respectively, then $t$ is also a RV and its distribution can be approximated by \cite{J:Farid07,J:Boluda-Ruiz16}
\begin{equation}\label{Eq:Pointing}
f_{t}(t) = \frac{\phi^{2}}{A^{\phi^{2}}} \, t^{\phi^{2}-1} \, , \, 0\leq t \leq A \, .
\end{equation}
The distribution parameters $\phi$ and $A$ are calculated from the receiver aperture radius $a$, the beam width at the receiver $w_{z}$ and the misalignment mean values and variances \cite{J:Boluda-Ruiz16}.

We now combine Eq.~\eqref{Eq:chisqSerrorprobPPMavg} and \eqref{Eq:Pointing} to find that the weight function equals
\begin{equation}\label{Eq:weightfun}
\begin{split}
w(n) &= \frac{\phi^{2}\, z_{q}^{n}}{A^{\phi^{2}}\,n!} \, \int_{0}^{A} t^{n+\phi^{2}-1} \, \exp \left( - z_{q} \, t \right)\, dt \\
&= \frac{\phi^{2}}{(A\,z_{q})^{\phi^{2}}\,n!} \, \gamma \left( n+\phi^{2}, A \, z_{q} \right) \, ,
\end{split}
\end{equation}
where $\gamma(\cdot)$ is the lower incomplete Gamma function. Equation~\eqref{Eq:weightfun} is valid as long as one of the variances is non-zero and simplifies to
\begin{equation}\label{Eq:weightfunlim}
w(n) = \frac{(A\,z_{q})^{n}}{n!} \, \exp \left( -A\,z_{q} \right)
\end{equation}
in a static misalignment scenario where both variances are equal to zero.

The ABEP is plotted in Fig.~\ref{fig:ABEP} for a modulation order equal to $Q=16$, two systems with noise modes equal to $M=2$ and $M=200$ \cite{J:Stevens12}, and varying beam widths. Fig.~\ref{fig:ABEP}(a) shows the ABEP for a system with a fixed misalignment in the $x-$direction ($\mu_{x} = 10\,a,\,\mu_{y} = 0$ and $\sigma_{x} = \sigma_{y} = 0$). In this static scenario, the ABEP is minimized for $w_{z}\simeq15\,a$, which corresponds to the optimal overlap between the beam and the aperture, and this is observed for both noise modes under consideration. Fig.~\ref{fig:ABEP}(b) shows the ABEP for a system with the same misalignment and non-zero jitter ($\mu_{x}=10\,a,\,\mu_{y}=0$ and $\sigma_{x}=\sigma_{y}=a$). The introduction of jitter modifies the ABEP behavior and the $w_{z} \simeq 15\,a$ beam provides the optimal performance for an ABEP up to approximately 10\textsuperscript{-6} in the $M=2$ system. However, the beam width needs to be increased to $w_{z} \simeq 20 \, a$ as more power becomes available and the ABEP improves. Regarding the $M=200$ system, the observations are similar but the beam width is required to increase at a lower ABEP equal to 10\textsuperscript{-5}.

\section{Beam Width Optimisation}
The previous discussion suggests that the ABEP system performance is significantly affected by the beam width size. The utilisation of a sub-optimal beam width is not too pronounced in a static misalignment scenario and the power penalty from using the wrong width is fixed and amounts to a couple of dBs for all $E_{b}/N_{0}$. However, in non-static scenarios the power penalty that is introduced from utilising narrower than the optimal beams can become very significant at increased powers. As a result, an optimisation of the beam width is required and this is performed by repeatedly solving Eq. \eqref{Eq:chisqSerrorprobPPM} and \eqref{Eq:weightfun}. To this end, we first solve the equations for a relative accuracy of $\delta w_{z}=\pm \, a$ so as to locate a coarse ABEP minimum. We then refine the solution by reducing the relative accuracy to $\delta w_{z}=\pm 0.1 \, a$. Despite the fact that several optimizations steps are required to locate the ABEP minimum for each $E_{b}/N_{0}$, especially at high energies, Eq. \eqref{Eq:chisqSerrorprobPPM} and \eqref{Eq:weightfun} are computationally efficient and can be evaluated using standard computer hardware within acceptable time.

\begin{figure*}[!t]
\centering
\subfloat{\includegraphics[keepaspectratio,width=0.9\columnwidth]{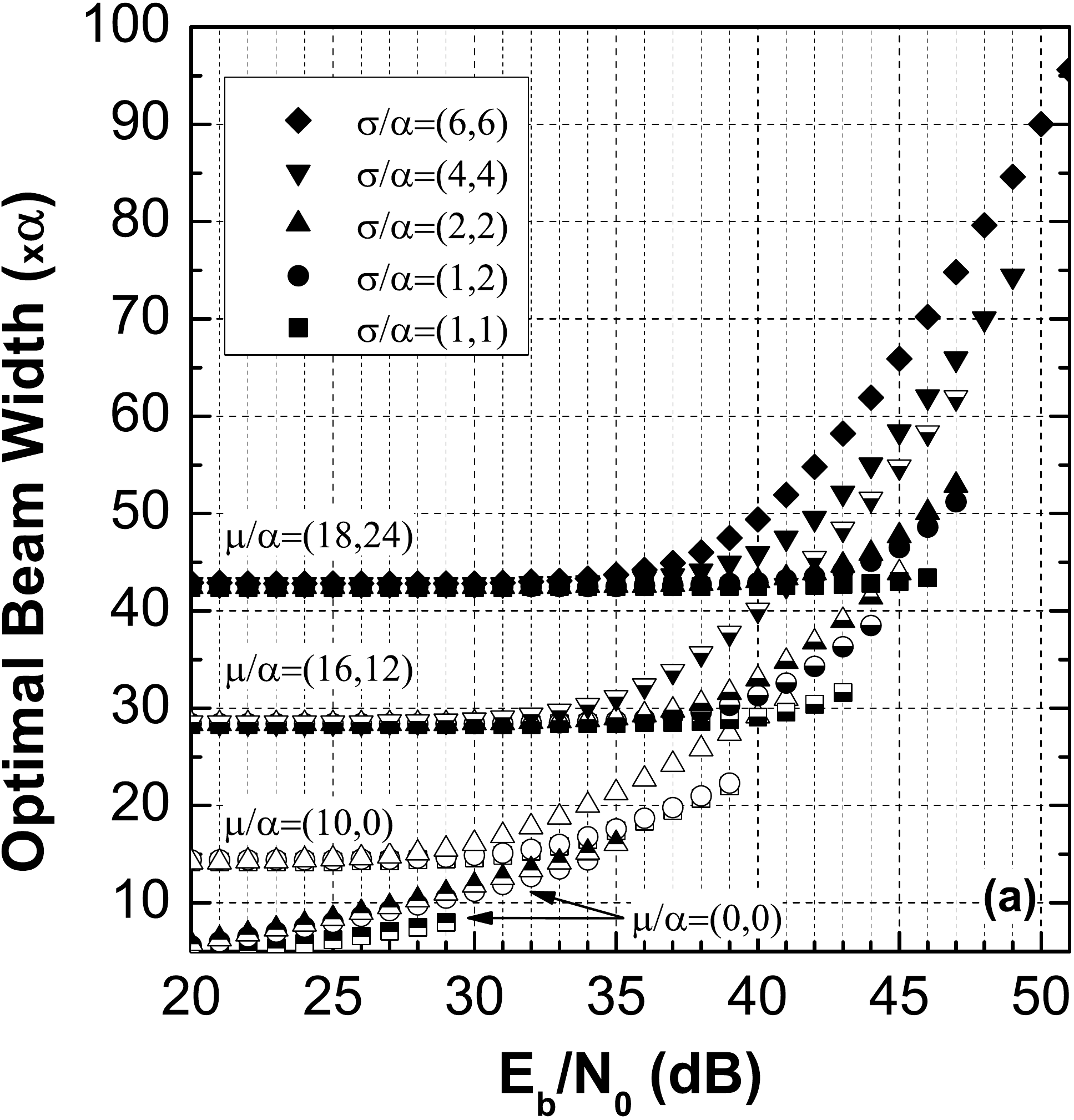}
\label{Fig:RoptQ16}}
\hfil
\subfloat{\includegraphics[keepaspectratio,width=0.9\columnwidth]{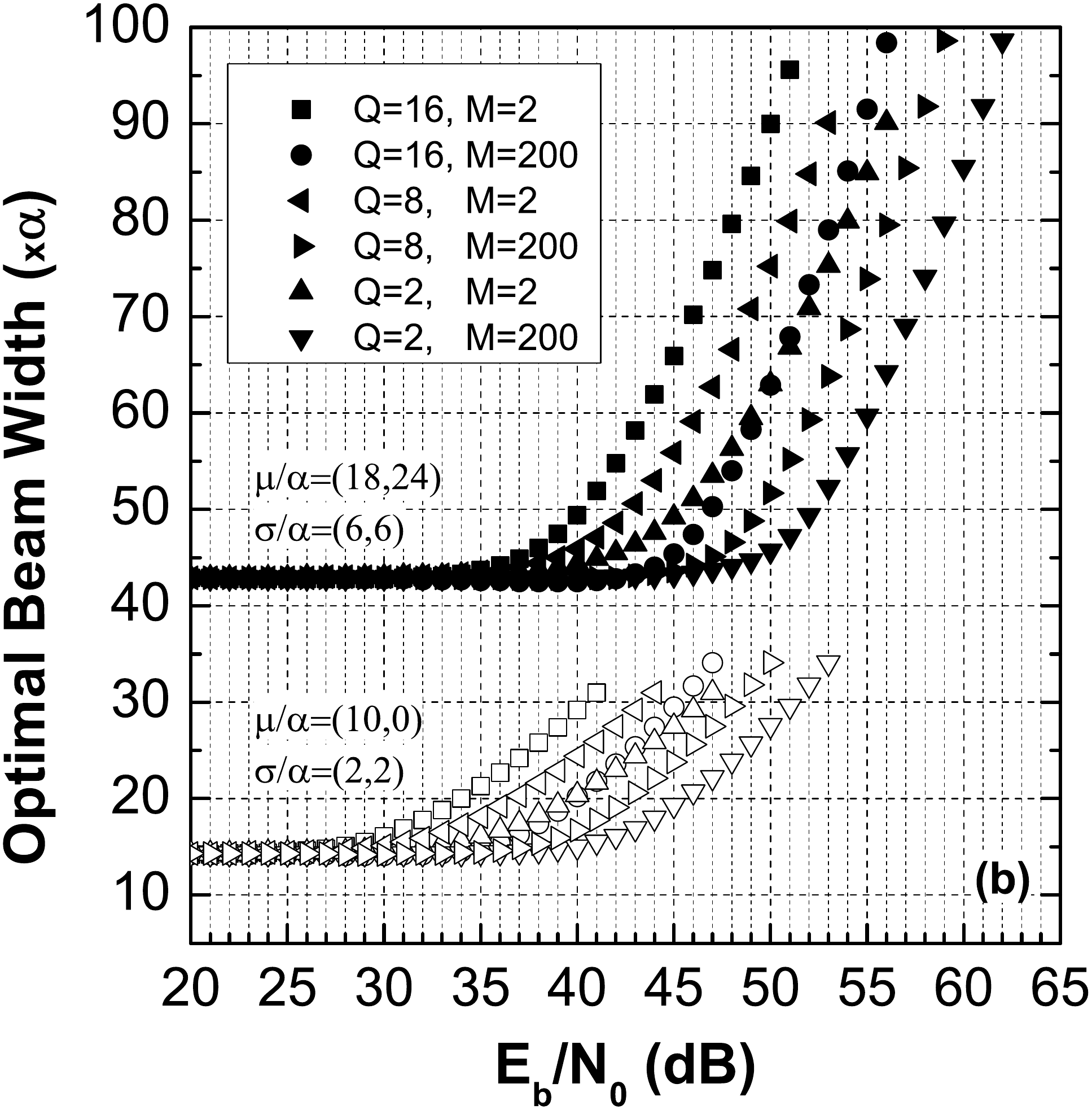}
\label{Fig:RoptQM}}
\caption{Optimal beam width dependence on (a) misalignment, and (b) modulation order and noise modes.}
\label{fig:Ropt}
\end{figure*}

Fig.~\ref{fig:Ropt}(a) presents the optimal beam width for a $Q=16$ and $M=2$ system and for a number of possible pointing error scenarios, ranging from perfect alignment and limited jitter to significant misalignment and jitter. At low beam energies, the results show that the optimal width is almost constant and that its value equals approximately 1.5 times the average misalignment, which is consistent with what is observed in Fig.~\ref{fig:ABEP}(a). The misalignment jitter does not affect the optimal width at the lower end of the energy spectrum. At higher energies, the jitter has a more pronounced impact and broader beam widths are required in arrangements with more jitter, as expected.

Fig.~\ref{fig:Ropt}(b) presents the dependence of the beam width on the modulation order and the amplification noise modes. The constant lower energy value is not affected by either of these parameters, however both the modulation order and the noise modes affect the higher energy response. Moreover, higher order modulation schemes are more efficient in terms of energy and allow for broader beams at any given energy level. On the other hand, additional energy is required on average when the noise modes are increased and the corresponding beams become narrower. For example, the energy gain that is obtained by increasing the modulation order from $Q=2$ to $Q=16$ is almost completely lost by increasing the noise modes from $M=2$ to $M=200$ and, as a result, the $Q=16$, $M=200$ system requires a comparable beam width to its $Q=2$, $M=2$ counterpart.

\section{Conclusion}
We have presented analytical results on the impact of pointing errors on the ABEP performance of optically pre-amplified PPM optical-wireless communication receivers. The results show that the misalignment between the beam and the receiving aperture imparts a significant power penalty, especially when a random jitter component is present, and that the ABEP is minimised by properly adjusting the width of the received beam. At low signal energies, the optimal beam width is practically constant and determined by the static misalignment. At higher energies, the optimal width increases and the rate of the increase depends on the misalignment jitter, the PPM modulation order and the noise modes. Finally, higher modulation orders and lower noise modes lead to more energy efficient systems and the utilisation of broader beams at any given energy level. This work extends the analytical framework that we have previously presented for the treatment of fading and can be enhanced further to include any combination of random effects with a known probability distribution function.

\bibliographystyle{IEEEtran}
\bibliography{IEEEabrv,ms}

\end{document}